\documentclass[
  ,final            
  ]
  {aipproc}

\layoutstyle{6x9}

\begin{document}

\title{Luminosity determination for the quasi-free nuclear reactions}


\author{P.~Moskal}{
  address={Institute of Physics, Jagellonian University, 30-059 Cracow, Poland \&
   \mbox{Institut f\"ur Kernphysik, Forschungszentrum J\"ulich, 52425 J\"ulich, Germany}}
}

\author{R.~Czy{\.z}ykiewicz for the COSY-11 collaboration}{
  address={Institute of Physics, Jagellonian University, 30-059 Cracow, Poland \&
   \mbox{Institut f\"ur Kernphysik, Forschungszentrum J\"ulich, 52425 J\"ulich, Germany}}
}

\begin{abstract}
A method for the calculation of the
luminosity for the proton-nucleus collisions based on the quasi-free proton-proton scattering
is presented. 
As an example of application the integrated luminosity for  the 
scattering of protons off the deuteron target is determined for the 
experiment of the quasi-free $pn\to pn\eta$ reaction performed by means of the COSY-11 facility.
\end{abstract}

\maketitle

\section{Introduction}

Due to the lack of the
pure neutron targets
a meson production in the proton-neutron reactions is usually realised e.g.
via the proton scattering off the deuteron.
This method takes advantage 
of the fact that the neutron's binding energy inside the deuteron is 
negligible relative to the kinetic energy of proton beams  needed to produce the meson in the nucleon-nucleon collisions~\cite{review,meson}.
In the analysis of such reactions it is assumed that the proton is acting only as a spectator and that it is 
on its mass shell  at the moment of the collision.

In this contribution we present the evaluation of the luminosity in such kind of experiments 
by means of the measurement of the quasi-free proton-proton scattering.
This method permits to take into account automatically the shadowing effects of the spectator nucleon,
and allows for the determination of the luminosity with a relatively small 
normalization uncertainty  thanks to the availability of the precise
cross sections for the proton-proton elastic scattering 
determined by the 
EDDA collaboration~\cite{edda} 
up to the beam momentum of 3.5~GeV/c.

\section{Luminosity}

\begin{figure}
\includegraphics[width=0.5\textwidth]{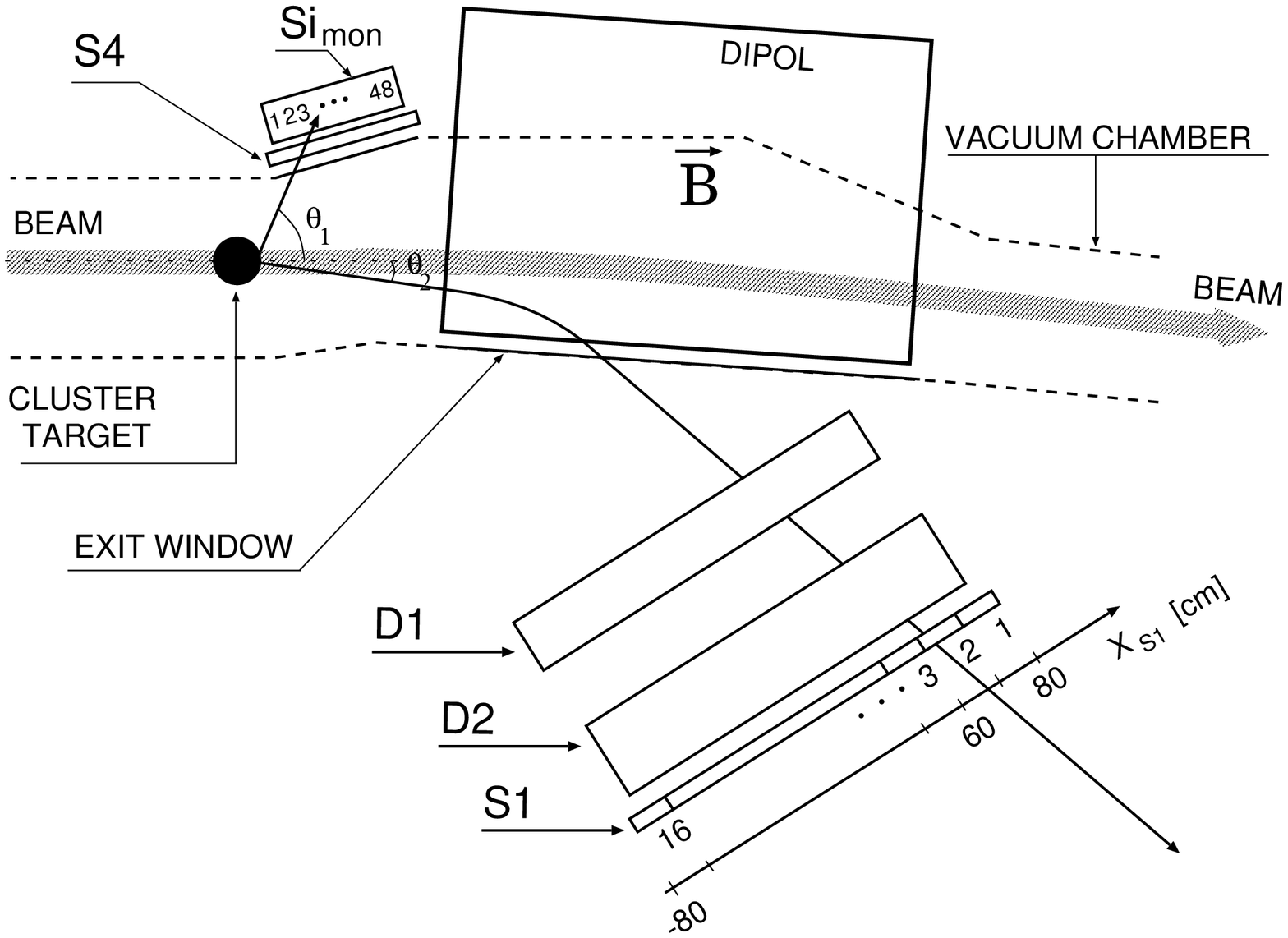}
\includegraphics[width=0.4\textwidth]{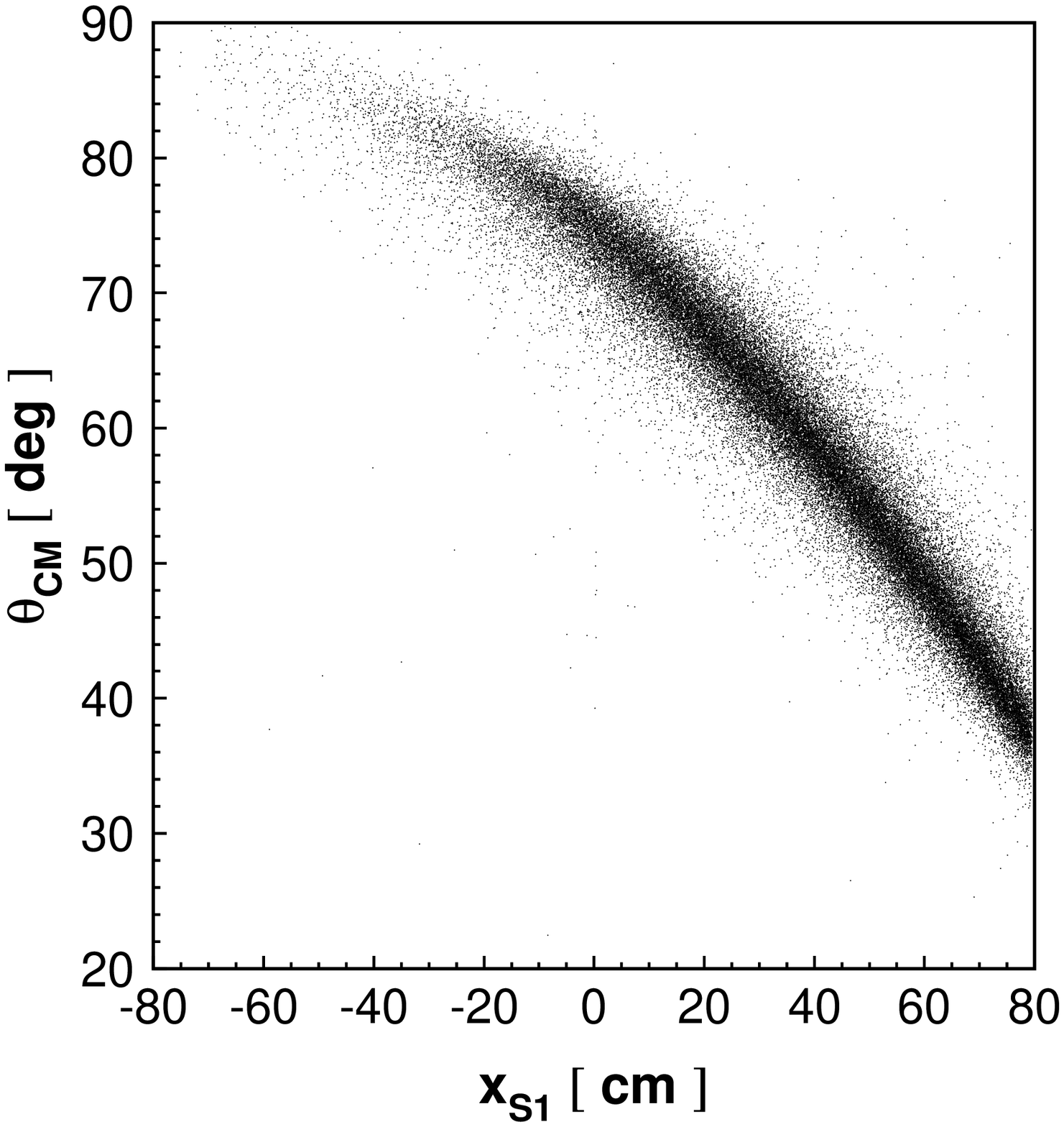}
\caption{
(left) Schematic view of the COSY-11 detection setup. Only the detectors
used for the measurements of the quasi-free proton-proton
scattering are shown. D1 and D2 denotes the stack of two drift chambers.
Si$_{mon}$ is the granulated silicon detector. S1 and S4 stand for the
scintillator detectors.
(right) Relation between 
the center-of-mass scattering angle and the position in the S1 detector
for the quasi-free proton-proton scattering at p$_{beam}=2.075$~GeV/c as 
has been obtained in the Monte-Carlo simulations.  
\label{uklad}
}
\end{figure}

In order to determine the luminosity for the quasi-free 
proton-neutron reaction we measured
quasi-free $pp\to pp$ reaction by detecting in coincidence
both scattered protons.
Here we will describe the measurement performed by means of the COSY-11 apparatus~\cite{brauksiepe,dombrowski,smyrski,klaja}
shown schematically in Figure~\ref{uklad} (left).
The recoil proton gives signal in
the scintillator detector S4 and subsequently reaches the granulated
silicon detector Si$_{mon}$, while the forward scattered proton is registered by the stack of
drift chambers D1 and D2 and scintillator array S1.
For triggering of the elastic scattered events the coincidence between signals from
 the S1 and S4 scintillators was required.

\begin{figure}
  \includegraphics[width=4.4cm]{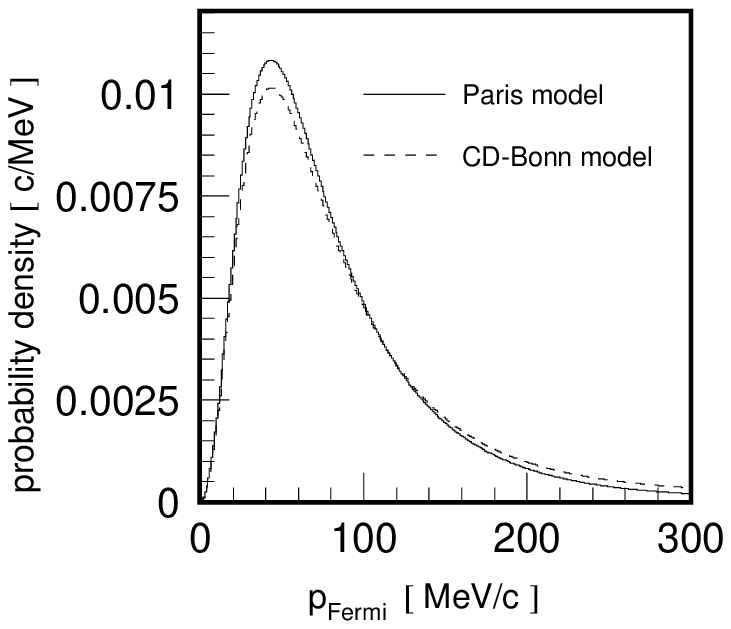}
  \includegraphics[width=4.7cm]{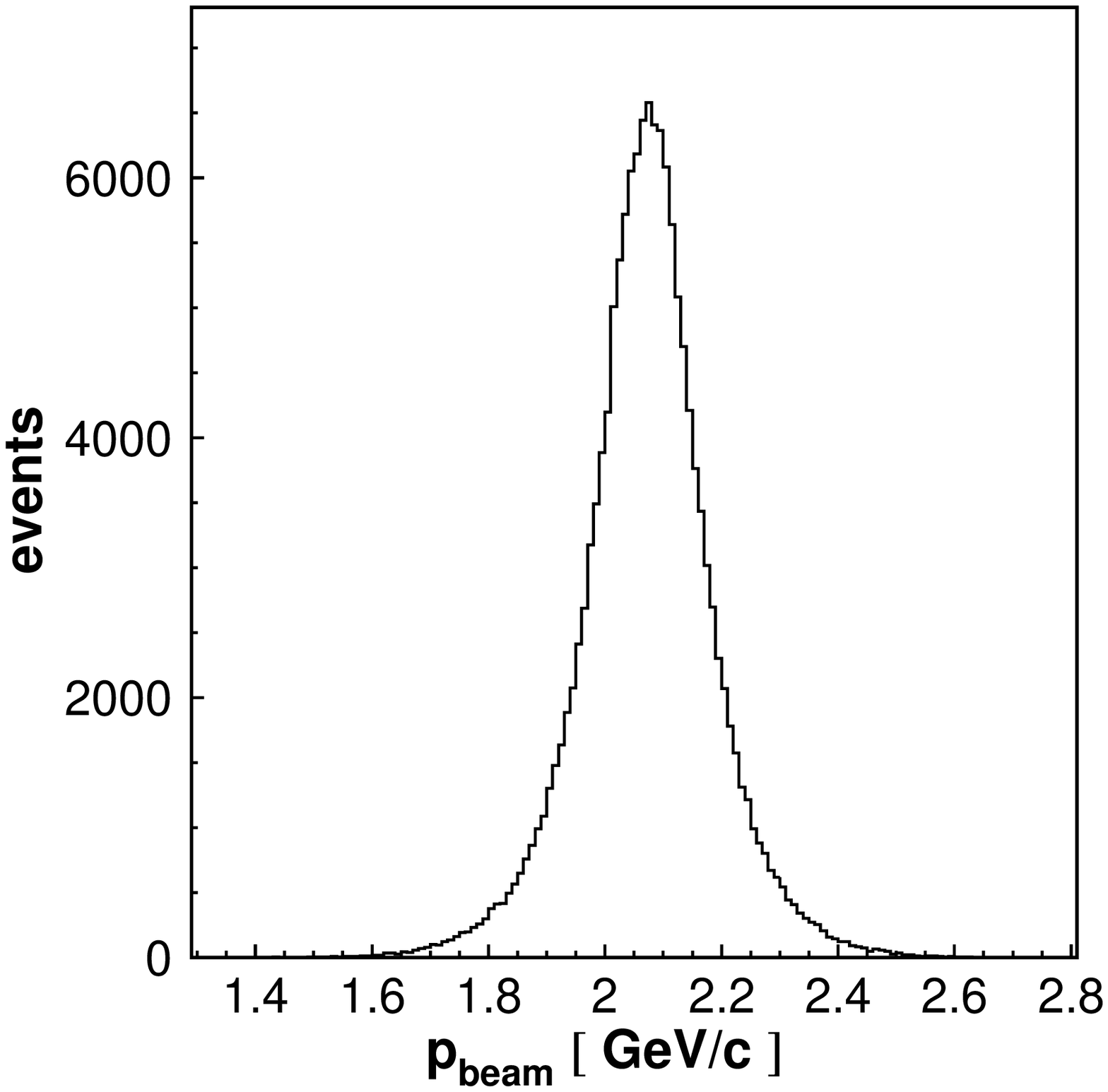}
  \includegraphics[width=4.8cm]{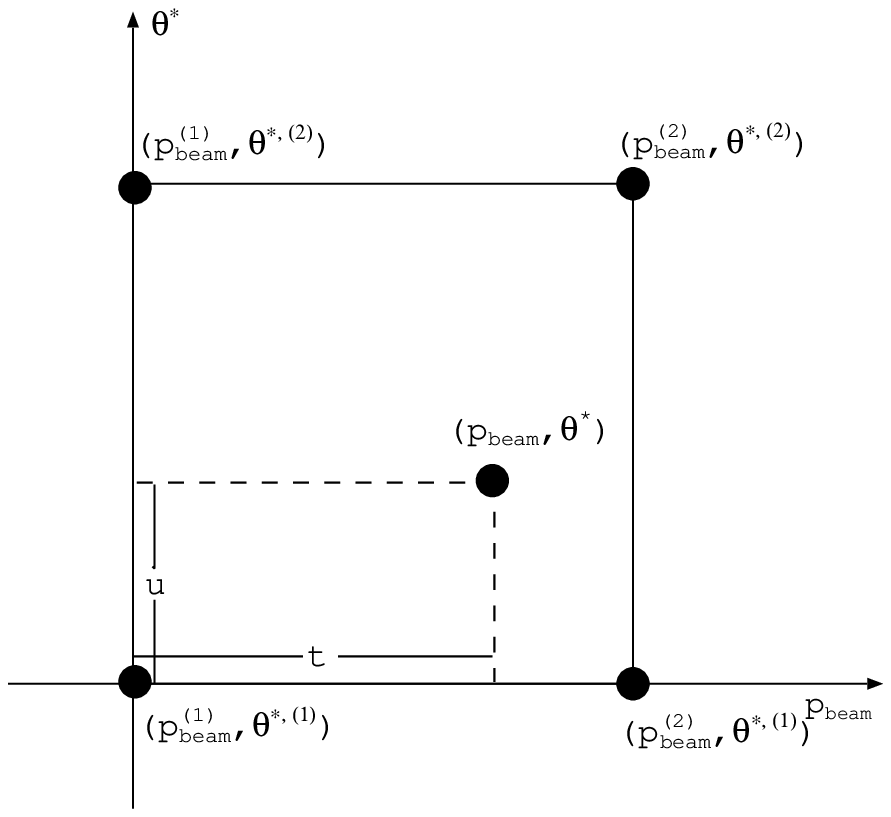}
  \caption{(left panel) Nucleon momentum distribution inside the deuteron
according to the Paris~\cite{paris} (full line) and CDBONN~\cite{cdbonn} (dotted line) potentials.
(middle panel) Distribution of the beam momentum as seen by the proton bounded inside a deuteron
hit by the beam proton with 
momentum
of $p_{beam}$~=~2.075~GeV/c.
(right)
 Bilinear interpolation of the differential cross
section $\frac{d\sigma}{d\Omega}(p_{beam},\theta^*)$.}
  \label{fermi}
\end{figure}

In the case of free proton-proton scattering 
the luminosity could be determined as a normalization 
constant between the measured angular distribution of the 
cross section and the 
corresponding spectrum known from the  previous experiments.
For the quasi-free reaction the evaluation becomes more complicated
due to the Fermi motion of nucleons inside the nucleus~(Fig.~\ref{fermi} (left)).
Since the direction and momentum of the bound nucleon may vary from
event to event this implies that 
the direction of the center-of-mass velocity of the colliding nucleons
as well as the total available energy for the reaction may also vary from event to event.
Therefore,  protons registered in the laboratory under a given scattering angle
or at a given part of the detection system, 
correspond to the finite 
range of scattering
angles in the proton-proton center-of-mass frame (see Fig.~\ref{uklad} (right)).
This implies that  the  experimental angular distributions cannot be directly 
compared to the literature values, and instead an evaluation of the luminosity
requires  simulations taking into account the Fermi motion of the nucleons,
and the variations of differential cross sections for the elastic scattering as a function
of the scattering angle and energy.

For each simulated event we know the generated Fermi momentum of the nucleon,
as well as the scattering angle of protons in their center-of-mass system.
This permits us to assign to each event a weight corresponding to the 
differential cross section, which is uniquely determined by the scattering angle
and the total collision energy $s$.

For the intuitive illustration of the size of the momentum spread caused by the Fermi motion
instead of the total center-of-mass energy we may equivalently consider the effective beam momentum
as it is seen from the nucleon inside the nucleus.  The distribution of the effective beam 
momentum depends on the value of the proton beam momentum
and as an example in Figure~\ref{fermi} (middle) we present it for the value of 2075~MeV/c 
as which used for the measurement
of the quasi free $pn\to pn\eta$ reaction~\cite{xx,zz}.
It is important to note that the distribution
of the equivalent beam momentum ranges from about 1.5~GeV/c up to
circa 2.5~GeV/c. In this momentum range the cross sections for the 
proton-proton elastic scattering vary significantly~\cite{edda} and therefore 
this effect cannot be neglected. 

In the following we will be more specific and will describe the derivation of the luminosity 
in terms of formulae exploited in the analysis.
For a free proton-proton scattering we could measure the number of events -- $\Delta N(\theta, \phi)$ scattered 
into the solid angle $\Delta\Omega(\theta, \phi)$ around the 
polar and azimuthal angles $\theta$ and $\phi$, respectively. In this case the angles in laboratory 
and in the center-of-mass systems are univocally related to each other. 
With the known differential cross section~\cite{edda} for proton-proton scattering into that
particular solid angle, and having known the value of the solid angle $\Delta\Omega(\theta, \phi)$ from the 
Monte-Carlo simulations the luminosity can be calculated
according to the formula:
\begin{equation}
L = \frac{\Delta N(\theta, \phi)}{\Delta\Omega(\theta, \phi) \frac{d\sigma}{d\Omega}(\theta, \phi)}.
\label{free}
\end{equation}

In the case of quasi-free proton-proton scattering  
the number of elastically scattered protons $\Delta N$ into a
solid angle $\Delta\Omega(\theta_{lab},\phi_{lab})$ is proportional
to $L$ -- the integrated luminosity over the time of measurement,
and also to the inner product of the differential cross section
for scattering into the solid angle around $\theta^*$ and $\phi^*$ angles --
$\frac{d\sigma}{d\Omega}(\theta^*,\phi^*,p_F,\theta_F,\phi_F)$ -- and the
probability density of the distribution of the Fermi momentum $f(p_F,\theta_F,\phi_F)$:
\begin{eqnarray}
\Delta N_{exp}(\Delta\Omega(\theta_{lab},\phi_{lab})) = \hfill \hspace{4cm} \
\nonumber
\\
L \int_{\Delta \Omega(\theta_{lab},\phi_{lab})}{\frac{d\sigma}{d\Omega}(\theta^*,\phi^*,p_F,\theta_F,\phi_F) f(p_F,\theta_F,\phi_F) dp_F dcos\theta_F d\phi_F d\phi^* dcos\theta^*}.
\label{iii}
\end{eqnarray}

The angles $\theta^*$ and $\phi^*$ are expressed in the
proton-proton center-of-mass system, while the angles $\theta_{lab}$ and $\phi_{lab}$
are considered in the laboratory system. 
In the case of the complex detection geometry with a magnetic field a solid angle corresponding
to a given part of the detector cannot in general be expressed in a closed analytical form.
Therefore, integral in Equation~\ref{iii} must be computed using the 
Monte-Carlo simulation programme, containing the exact geometry of the detection system and 
taking into account bending of particles trajectories in the magnetic field as well 
as  detection and reconstruction efficiencies.
For the evaluation of a given event by the Monte-Carlo programme first 
we choose randomly a  momentum of a nucleon inside a deuteron according to the Fermi distribution~\cite{paris} (Fig.~\ref{fermi} (left)).
Next the total energy $s$ for the proton-proton scattering and the vector of the center-of-mass velocity are determined.
Then, we generate isotropically a momentum of protons in the proton-proton center-of-mass frame.
Further on, according to the generated angle and the total-energy $s$ (or equivalently an effective beam momentum
seen by the struck nucleon) we assign to the event a probability equal to the 
differential cross section. Next, the momenta of protons are transformed to the laboratory frame
and are used as an input in the simulation of the detectors signals with the use of the GEANT computing package. 

The differential cross sections 
$\frac{d\sigma}{d\Omega}(\theta^*,\phi^*,s)$, with 
$s$ being dependent on $p_F,\theta_F,\phi_F$, and $p_{beam}$, and with
$\theta^*$ and $\phi^*$ denoting the scattering angles
in the proton-proton center-of-mass system were calculated 
using the cross section data base for the $pp\to pp$ reaction~\cite{edda}~\footnote{EDDA group 
has gathered the data for the excitation
functions $\frac{d\sigma}{d\Omega}(\theta^*,p_{beam})$ for the elastic $pp\to pp$ process
at 108 different proton kinetic energies, ranging from 240~MeV up to 2577~MeV.
In the measurements the center-of-mass scattering angles of protons ($\theta^*$)
from 30$^{\circ}$ up to 90$^{\circ}$ have been covered. Both the kinetic and angular
ranges are sufficient to cover our needs in calculating the corresponding differential
cross sections.}.
For this purpose we have applied a bilinear interpolation
in the momentum--scattering angle plane
and calculated differential cross section 
according to the formula:
\begin{eqnarray}
\frac{d\sigma}{d\Omega}(p_{beam},\theta^*) = (1-t)(1-u) \frac{d\sigma}{d\Omega}(p^1_{beam},\theta^{*,1}) + t(1-u) \frac{d\sigma}{d\Omega}(p^2_{beam},\theta^{*,1}) + 
\nonumber
\\
tu \frac{d\sigma}{d\Omega}(p^2_{beam},\theta^{*,2}) + (1-t)u \frac{d\sigma}{d\Omega}(p^1_{beam},\theta^{*,2}),
\label{bili}
\end{eqnarray}
where variables $t$ and $u$ are defined in Figure~\ref{fermi} (right).
These cross sections were used
as the weights 
of the elastically scattered events in the Monte-Carlo calculations.

$\Delta N_{exp}$ from Equation~\ref{iii} can be determined as a number of elastically scattered protons
registered in a given part of the detector system.
In order to calculate the integral on the right hand side of this equation we simulated $N_0$ events
according to the procedure described above. Due to the weights assigned to the events 
the integral is not dimensionless and its  units correspond to the units of the cross sections used for the
calculations.  
The number obtained from the Monte-Carlo simulations must be then normalized such that 
the integral over the full solid angle equals to the total cross section for the elastic scattering
averaged over the distribution of the total reaction energy $s$ resulting from the Fermi distribution 
of the target nucleon. In the absence of the Fermi motion it should be simply equal 
to a total elastic cross section
for a given beam momentum. This means that we need to divide the resultant integral by
the number of generated events $N_0$ and multiply it by the factor of $2\pi$.
A factor $2\pi$ comes from the normalization of the
differential cross section $\frac{d\sigma}{d\Omega}(p_{beam},\theta^*,\phi^*)$,
regarding the fact that 
protons taking part in the scattering are indistinguishable. 
Hence, the formula for the calculation of the integrated luminosity
for the quasi-free reaction reads:
\begin{equation}
L = \frac{N_0 \hspace{2mm}\Delta N_{exp}}{2\pi \int_{\Delta \Omega(\theta_{lab},\phi_{lab})}{\frac{d\sigma}{d\Omega}(\theta^*,\phi^*,p_F,\theta_F,\phi_F) f(p_F,\theta_F,\phi_F) dp_F dcos\theta_F d\phi_F d\phi^* dcos\theta^*}},
\label{final}
\end{equation}
where the normalization constant $N_{0}/2\pi$ is subject to the Monte-Carlo method used for the
integral computation.

\section{Example of application} 

We have applied the above described method 
for the evaluation of the luminosity for the measurement of the $pn\to pn\eta$ 
reaction with the COSY-11 facility~\cite{xx,zz,aip}.

\begin{figure}[h]
\includegraphics[width=4.6cm]{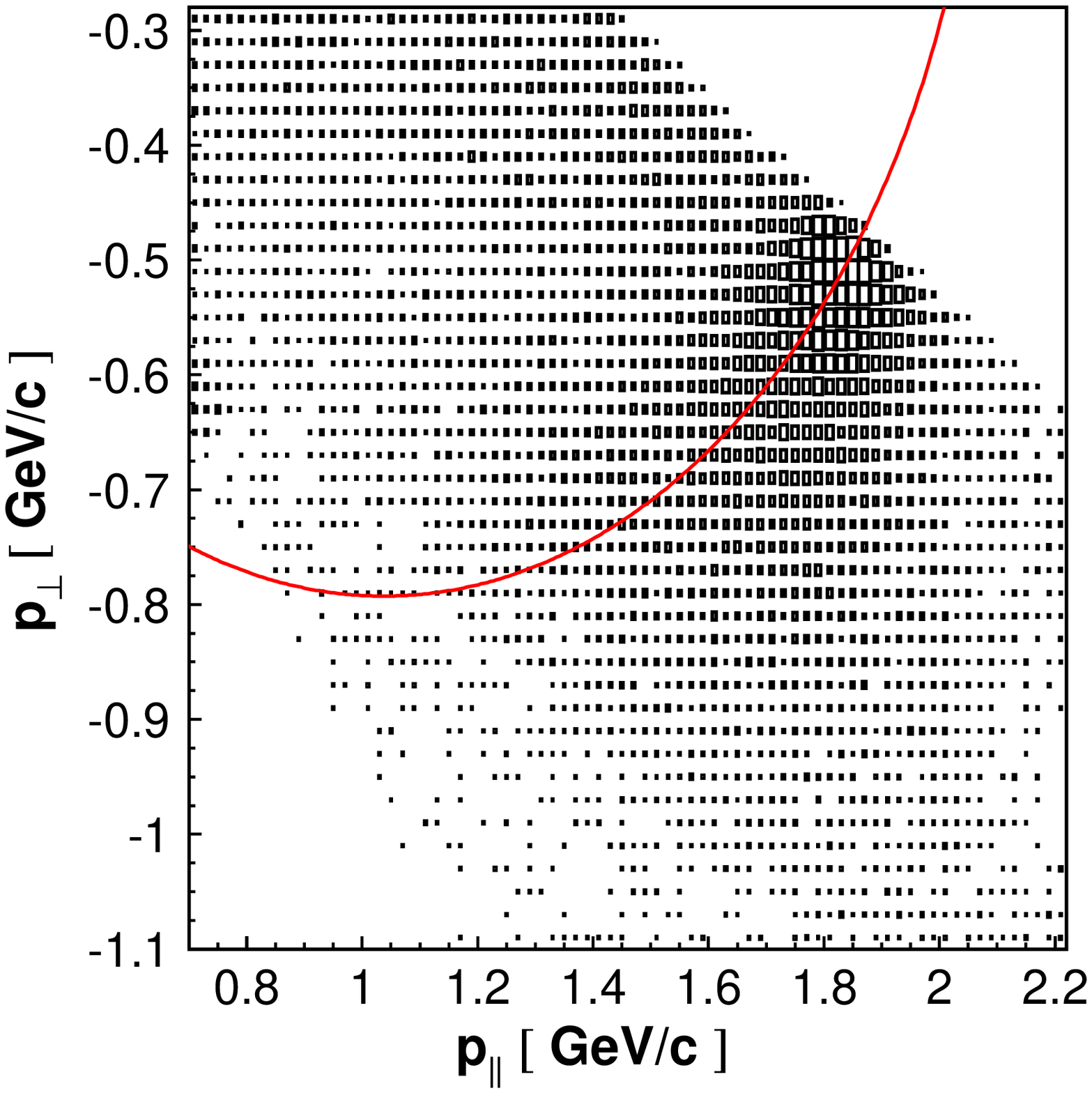}
\includegraphics[width=4.6cm]{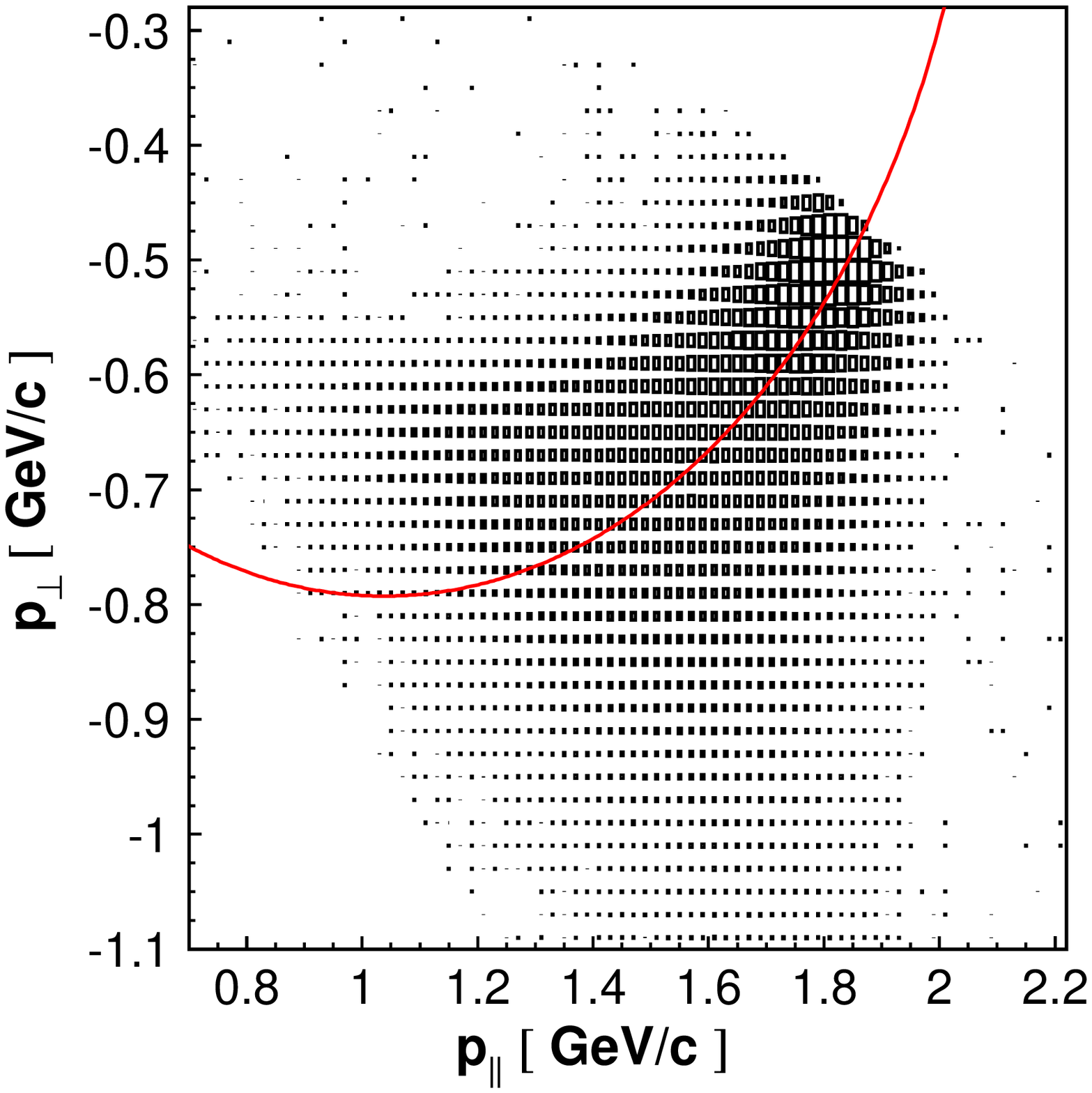}
\caption{Parallel versus transversal momentum component of the
reconstructed fast proton momentum as obtained in the experiment (left)
and in the simulations (right). Superimposed lines correspond to the expected
kinematical ellipses.
\label{elipsa}}
\end{figure}

Events corresponding to the elastically scattered protons have been identified
on the basis of the momentum distributions. 
The momentum of the fast scattered proton, whose trajectory
has been registered in the drift chambers, can be
reconstructed, and the transversal versus the parallel momentum component may be
plotted as it is done in left panel of Figure~\ref{elipsa}.
The signal from the elastic scattered protons appears as an clear enhancement around the expected
kinematical ellipse.

\begin{figure}[h]
\includegraphics[width=3.5cm]{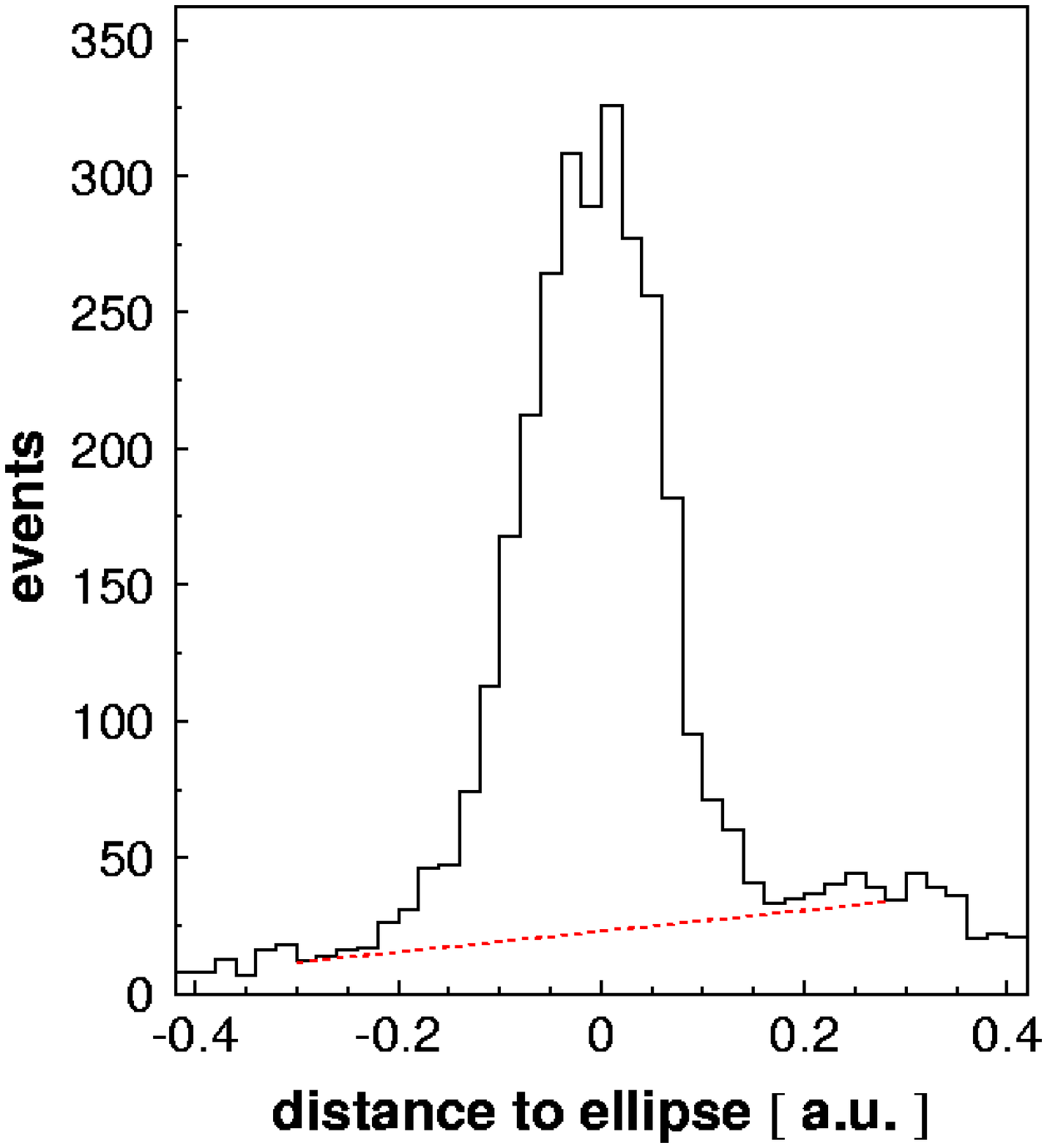}
\includegraphics[width=3.5cm]{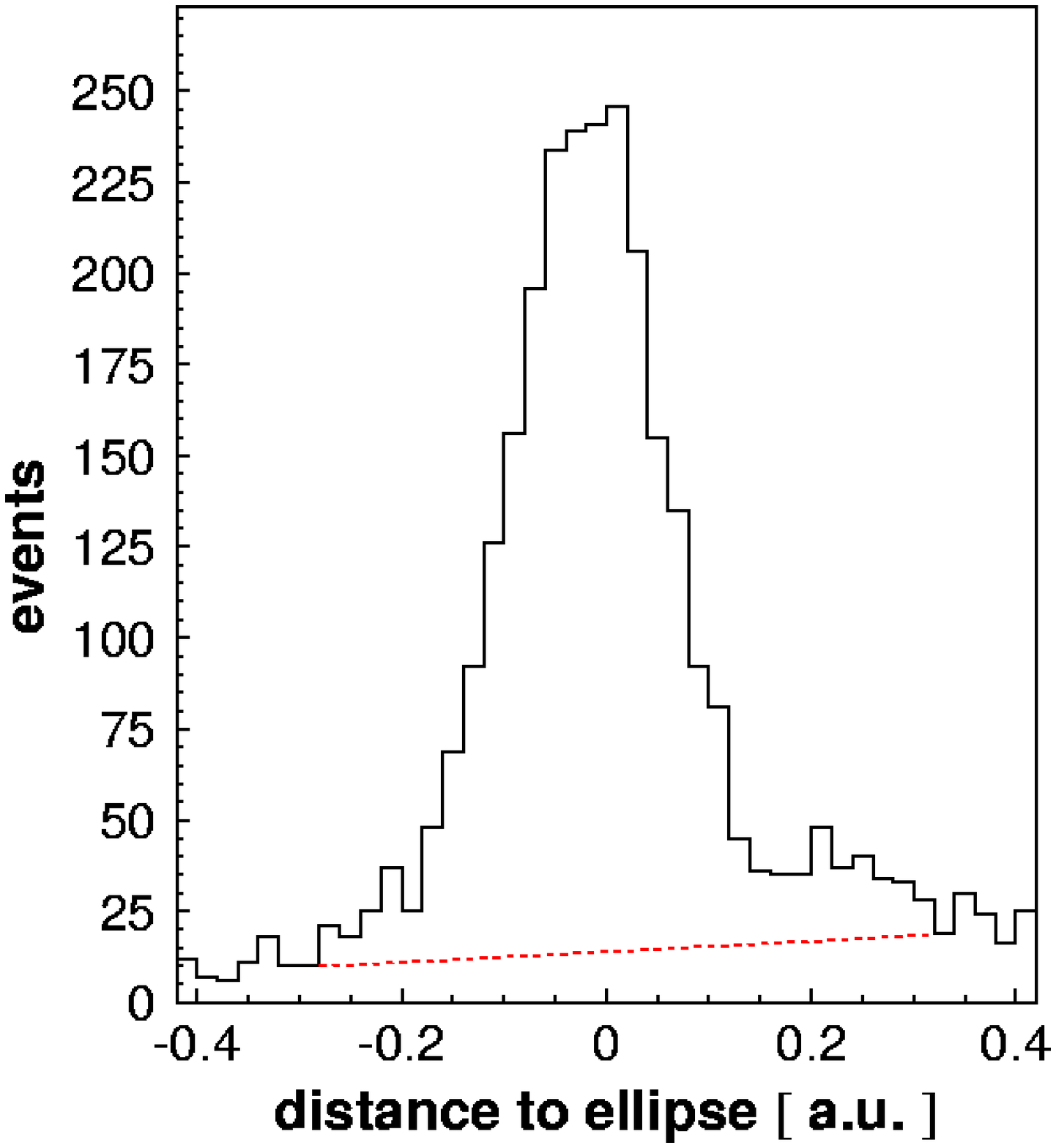}
\includegraphics[width=3.5cm]{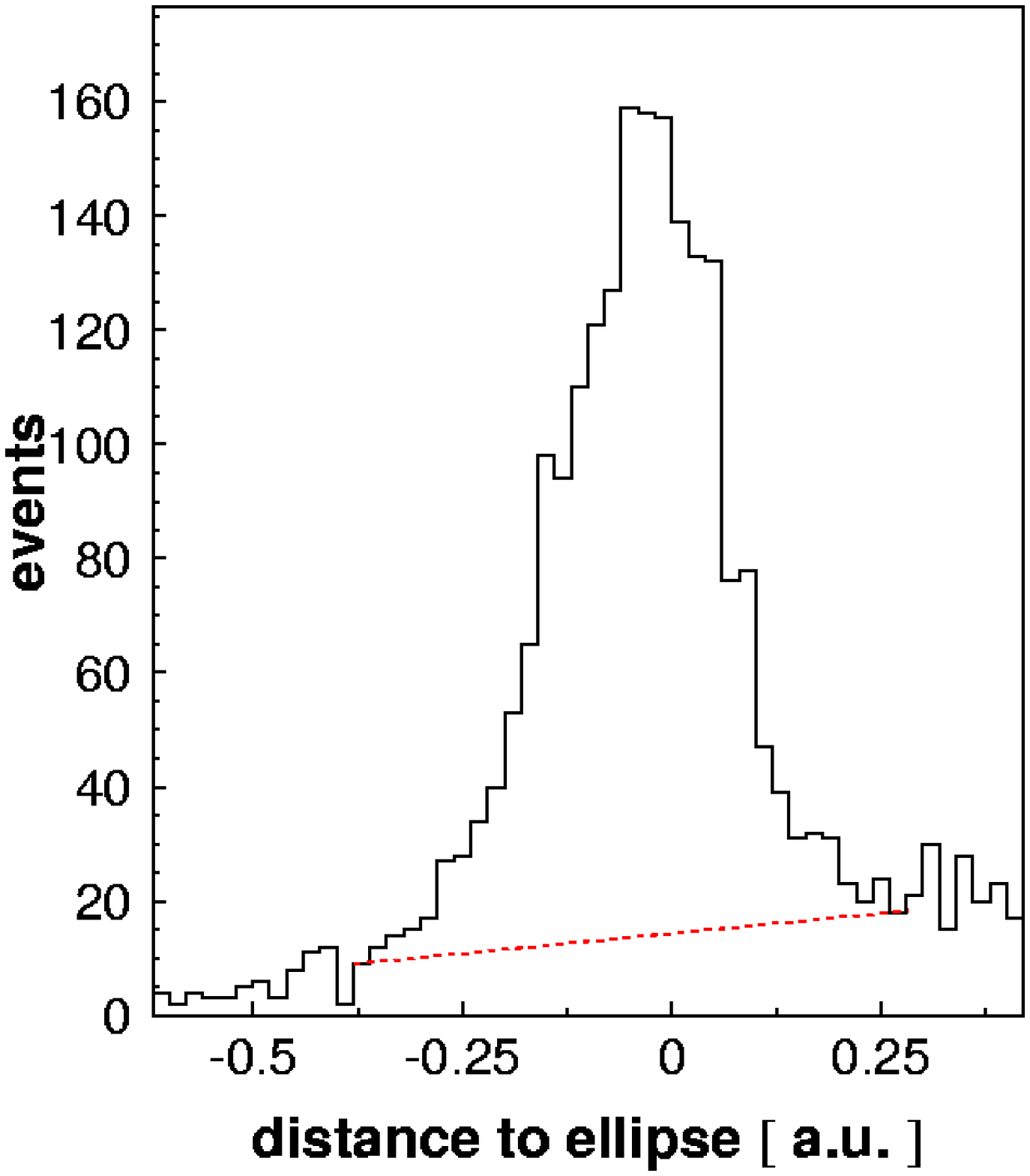}
\includegraphics[width=3.5cm]{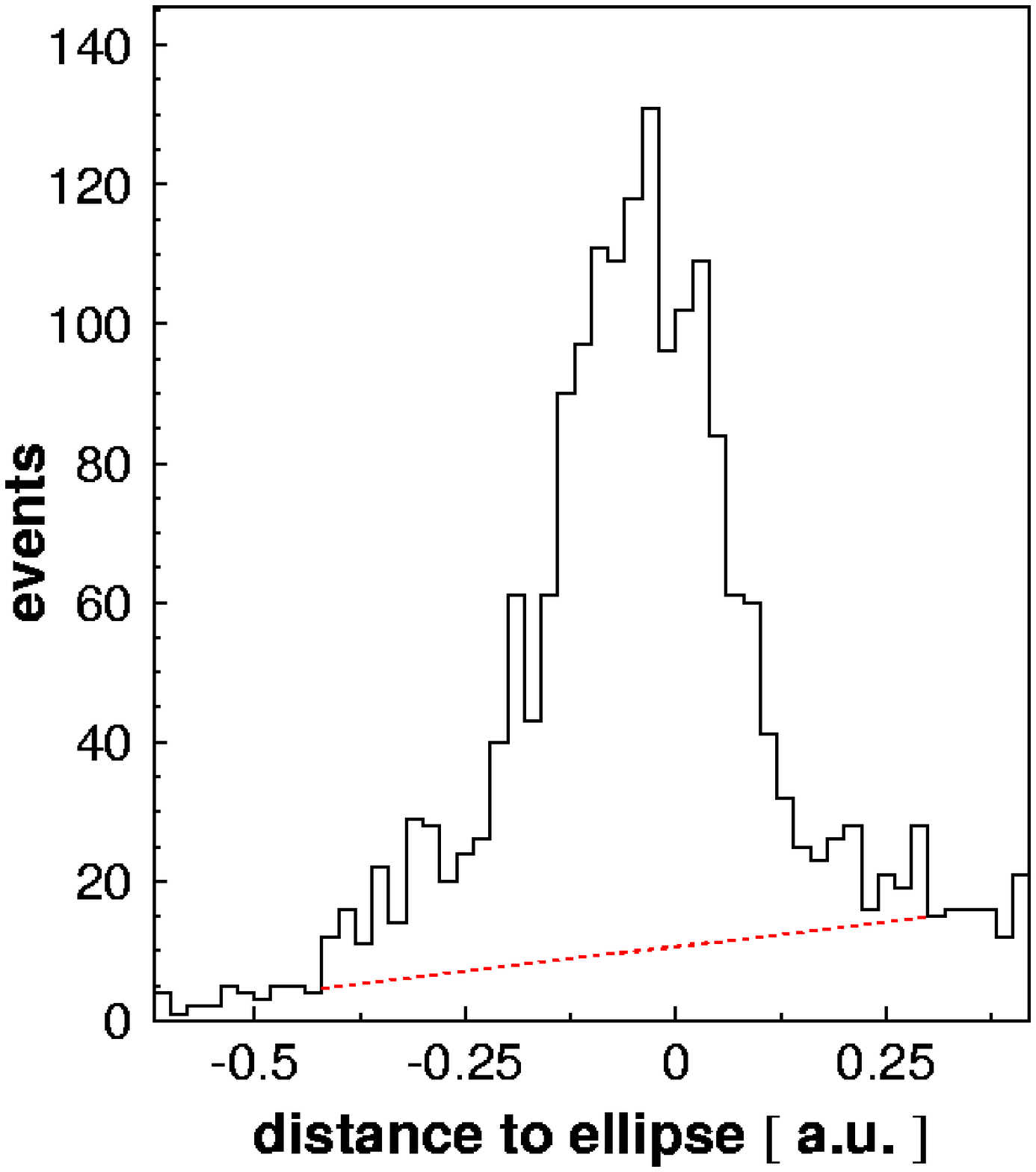}
\caption{Projection along the expected kinematical ellipse 
of the experimental event distribution 
from Figure~\ref{elipsa} (left) 
for four subranges of the S1 detector.
\label{exp_mc1}}
\end{figure}

For the calculations of the integrated luminosity,
the part of the S1 detector available for the elastically scattered protons 
was divided into four subranges.
In order to separate the background from the multi particle reactions
for each  subrange, the distribution  of the distance of the
points to the 
kinematical ellipse was determined. 
The result obtained for different sections of the S1 detector are presented in 
Figure~\ref{exp_mc1}.

\begin{figure}
\includegraphics[width=3.5cm]{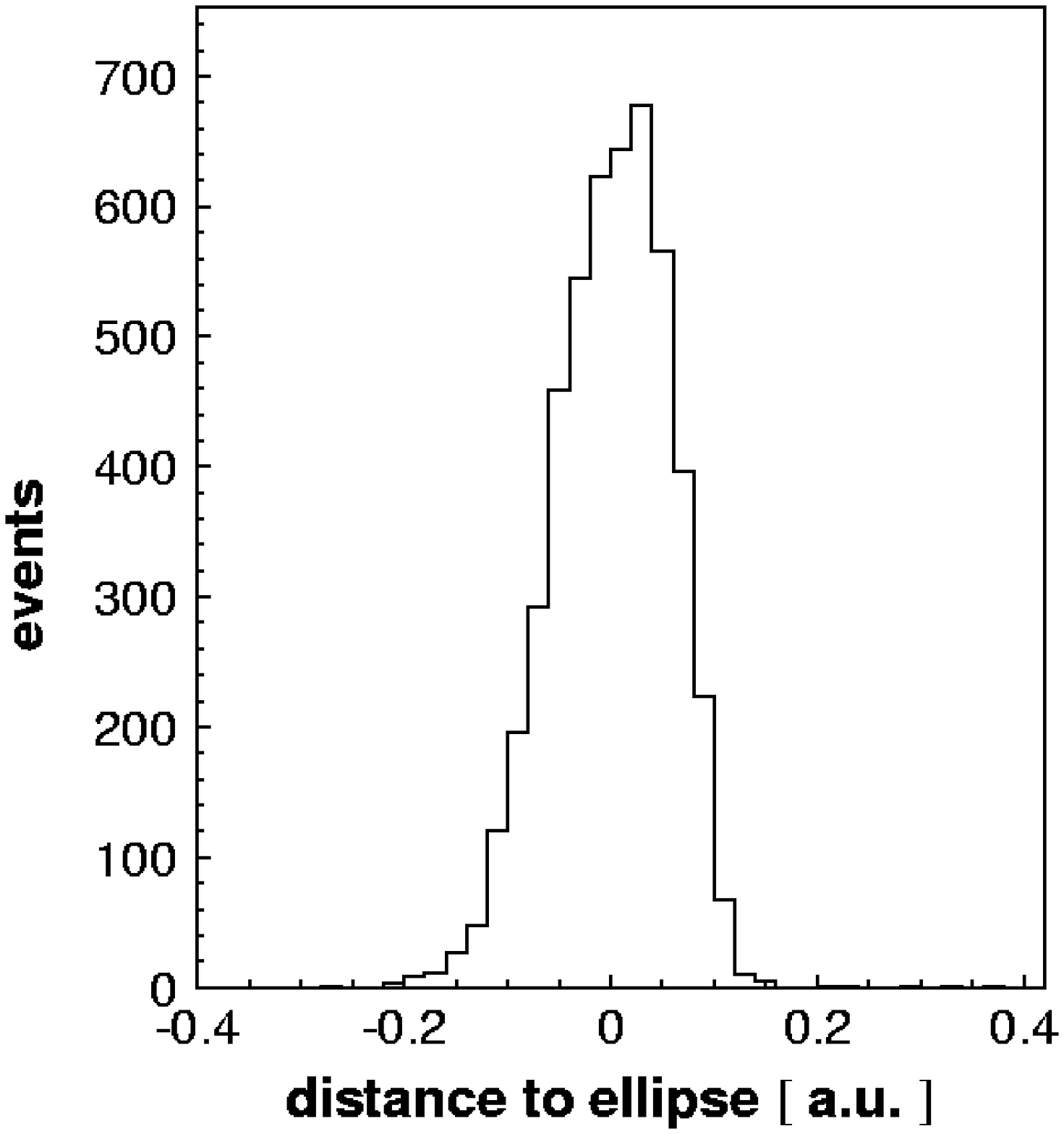}
\includegraphics[width=3.5cm]{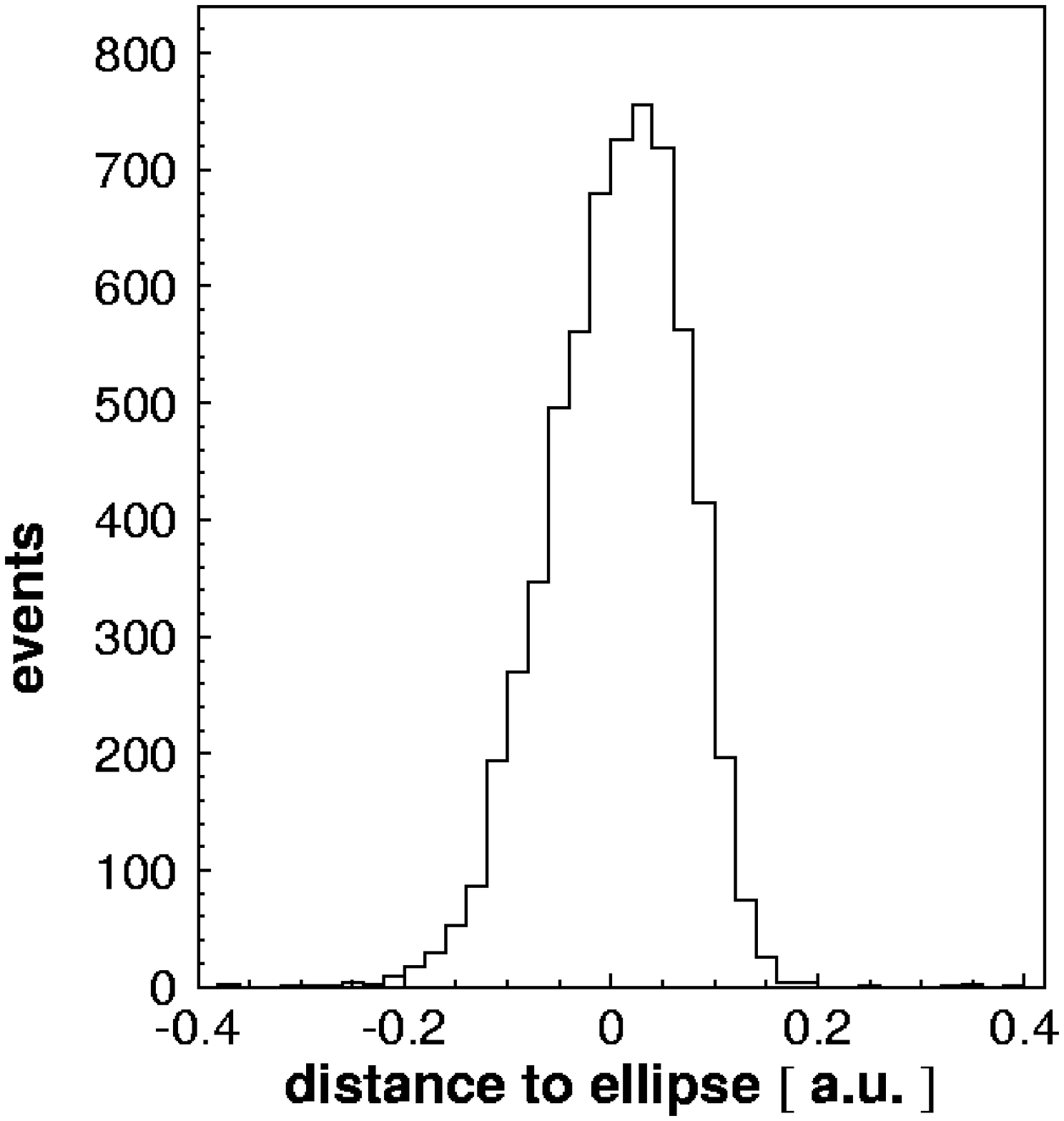}
\includegraphics[width=3.5cm]{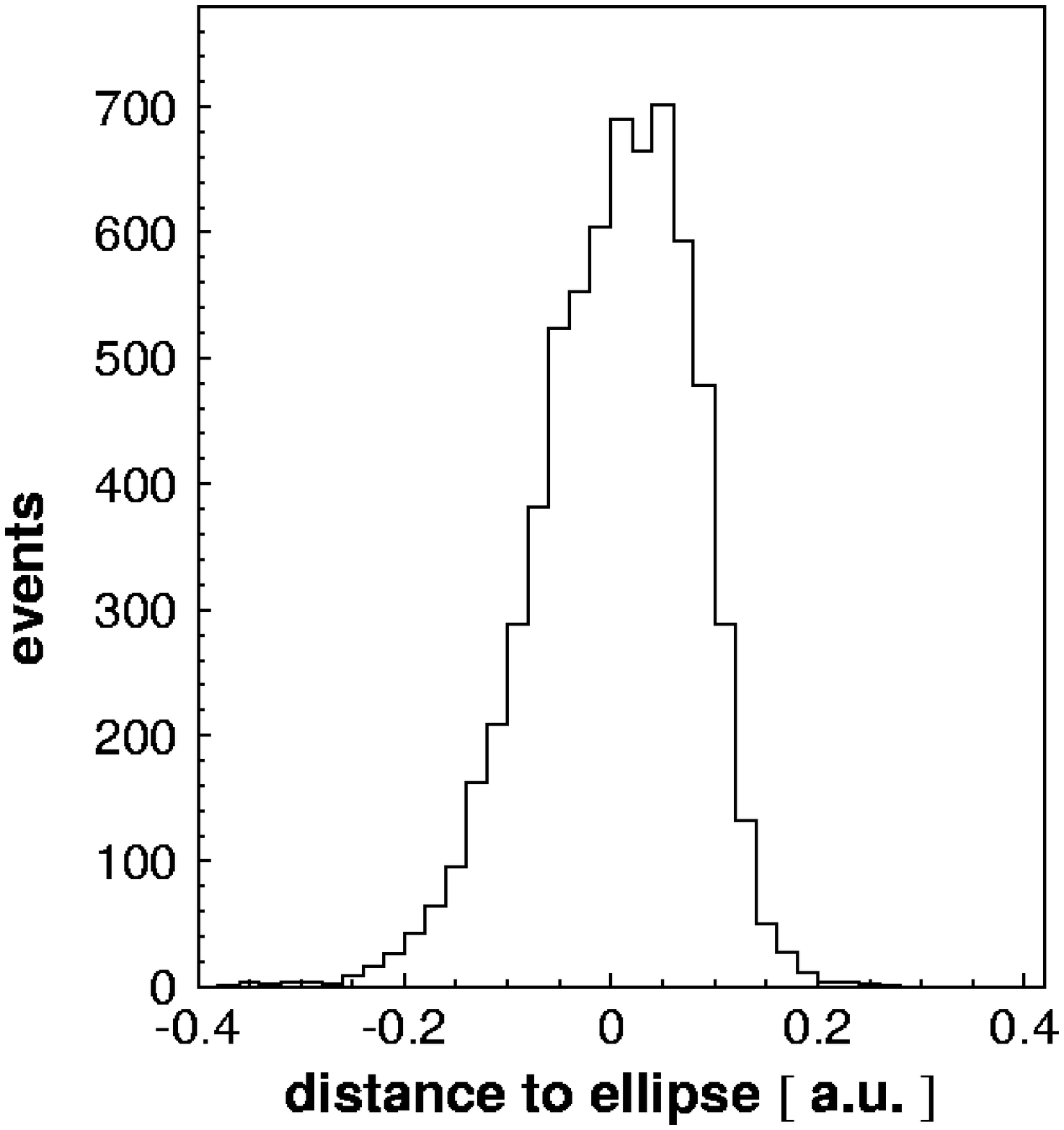}
\includegraphics[width=3.5cm]{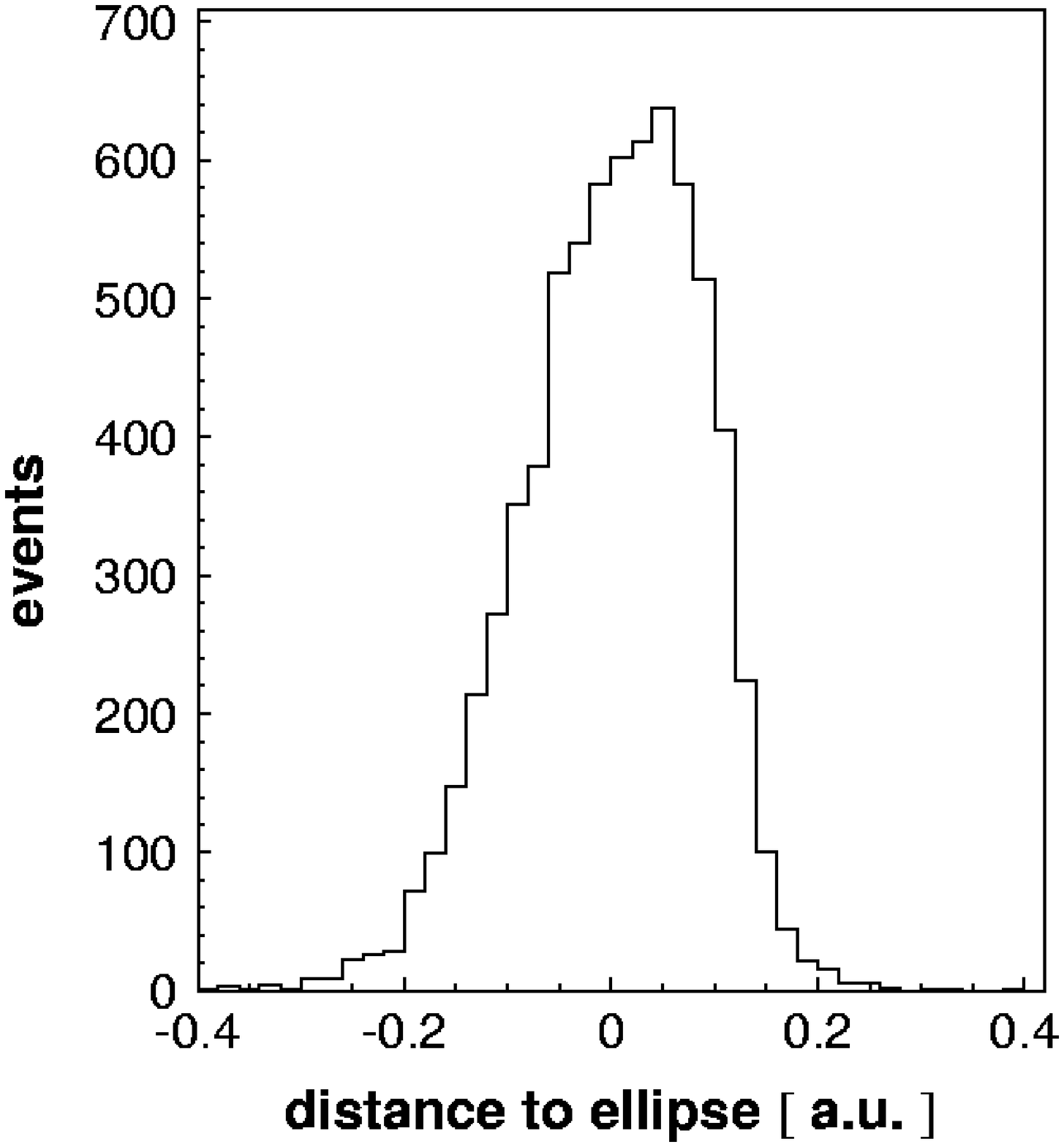}
\caption{Projection along the expected kinematical ellipse 
of the simulated event distribution 
from Figure~\ref{elipsa} (right) 
for four subranges of the S1 detector.
\label{exp_mc2}}
\end{figure}

A linear background cut~\footnote{The main source of the background
are the accidental coincidences originating from the production and
scattering processess.} has been performed and subsequently the
true scattering yields into a given range of the S1 detector
have been calculated.

For the determination of the integral of Equation~\ref{iii}
an N$_0=10^7$ quasi free $pp\to pp$
events have been simulated, and the response of the detectors has been generated
using the GEANT-3 code based simulation programme, maintaining
the experimental conditions of beam and target~\cite{moskal-nim}. Subsequently, the simulated events
have been analyzed in the same way as the experimental data.
Figure~\ref{exp_mc2} shows the simulated spectra analogous to 
the experimental distributions of Figure~\ref{exp_mc1}.
As described in the previous sections each entry in the shown histograms was weighted according to the
differential cross sections and hence the integral of these spectra normalized to number of simulated events
and multiplied by a factor of $2\pi$ can be substituted for an integral in Equation~\ref{iii}.
Thus the integrals of experimental (Fig.~\ref{exp_mc1}) and simulated histograms (Fig.~\ref{exp_mc2})
applied in Equation~\ref{iii} permitted to determine the luminosity for each of
the subrange of S1 detector separately.
The weighted average over the four quoted values of the integrated luminosity equals to
L~=~(2.08$\pm$ 0.03)$\cdot 10^{35}$ cm$^{-2}$.

One source of the systematic error may be attached to the
background misidentification. 
Since the line describing the background can be well defined
on both: the left and the right side of the scattering peaks,
we assumed, conservatively, that the systematic
error due to the assumption of the linearity of the background in the way as
presented in Figure~\ref{exp_mc1} is less than 20\%. The background events 
constitute around 15\% of all events, hence the
overall systematic error due to the background subtraction is not greater
than 3\%.

Another source of the systematic error originates from the assumption of the
bilinear approximation of the cross section shown in Figure~\ref{fermi} (right).
To estimate this systematic uncertainty we made the zeroth-order
assumption in which 
instead of the interpolation we took the cross section value from the closest data point
in the effective proton beam momentum--scattering angle plane.
Higher order approximations of the differential
cross section given by Equation~\ref{bili} should be not greater than the
difference between the zeroth order approximation explained above and
the bilinear approximation. 
The  performed calculations shows that
this difference is smaller than 0.2\%.

Taking into account the normalisation error of the EDDA differential cross 
sections (equal to circa 4\%~\cite{edda,edda_error}), the systematical error 
originating from the assumption of the potential model of the nucleon bound
inside the deuteron (equal to about 2\%), and the two abovementioned sources of the 
systematical errors, we estimated the overall systematical 
error of the integrated luminosity to be not greater than 9.2\%.

\begin{theacknowledgments}
We acknowledge the support of the
European Community-Research Infrastructure Activity
under the FP6 programme (Hadron Physics, N4:EtaMesonNet,
RII3-CT-2004-506078), the support
of the Polish Ministry of Science and Higher Education under the grants
No. PB1060/P03/2004/26, 3240/H03/2006/31  and 1202/DFG/2007/03,
and  the support of the German Research Foundation (DFG)
under the grant No. GZ: 436 POL 113/117/0-1.
\end{theacknowledgments}

\end{document}